\documentclass[useAMS]{emulateapj}
\usepackage{natbib,graphicx,subfigure,epstopdf}
\usepackage{amsmath, amssymb}
\slugcomment{\sc Accepted for Publication in ApJ Letters}
\shorttitle{Lack of Obscuring Tori in BL Lacs}
\shortauthors{Plotkin et al.}

\defcitealias{plotkin10}{P10}

\begin{document}

\title{The Lack of Torus Emission from BL Lacertae Objects: An Infrared View of Unification with WISE}
\shorttitle{Lack of Torus Emission from BL Lacs}

\author{
Richard.~M.~Plotkin,\altaffilmark{1} 
Scott~F.~Anderson,\altaffilmark{2} 
W.~N.~Brandt,\altaffilmark{3,4}
Sera~Markoff,\altaffilmark{1}
Ohad~Shemmer,\altaffilmark{5}
Jianfeng~Wu\altaffilmark{3,4}
}

\altaffiltext{1}{Astronomical Institute `Anton Pannekoek', University of Amsterdam, Science Park 904, 1098 XH, Amsterdam, the Netherlands; r.m.plotkin@uva.nl}
\altaffiltext{2}{Department of Astronomy, University of Washington, Box 351580, Seattle, WA 98195, USA}
\altaffiltext{3}{Department of Astronomy and Astrophysics, Pennsylvania Sate University, 525 Davey Laboratory, University Park, PA 16802, USA}
\altaffiltext{4}{Institute for Gravitation and the Cosmos, Pennsylvania State University, University Park, PA 16802, USA}
\altaffiltext{5}{Department of Physics, University of North Texas, Denton, TX 76203, USA}

\begin{abstract}
 We use data from the \textit{Wide-Field Infrared Survey Explorer} (\textit{WISE}) to perform a statistical study on  the mid-infrared (IR) properties of a large number ($\sim$10$^2$) of BL~Lac objects --- low-luminosity Active Galactic Nuclei (AGN) with a jet beamed toward the Earth.  As expected, many BL~Lac objects are so highly beamed that their jet synchrotron emission dominates their IR spectral energy distributions.  In other BL~Lac objects, however, the jet is not strong enough to completely dilute the rest of the AGN emission.  We do not see observational signatures of the dusty torus from these weakly beamed BL~Lac objects.  The lack of observable torus emission is consistent with  suggestions that BL~Lac objects are fed by radiatively inefficient accretion disks.  Implications for the ``nature vs.\ nurture" debate for FR~I and FR~II radio galaxies are briefly discussed.  Our study supports the notion that, beyond orientation,  accretion rate plays an important role in AGN unification.

\end{abstract}


\section{Introduction}
\label{sec:intro}

The standard orientation-based unification paradigm posits that every Active Galactic Nucleus (AGN) is comprised of the same basic components: an accretion disk around a supermassive black hole, a broad emission line region (BELR), an obscuring torus, and a narrow line region; radio-loud AGN also launch large-scale relativistic jets \citep[e.g.,][]{antonucci93,urry95}.   The torus is a key  component in this unified model, as it can block the BELR along certain lines of sight, and its dust  also reprocesses UV/X-ray radiation into the infrared (IR).   In simple unification, the only other way to hide the BELR is also via a geometric argument:  emission lines are outshown by a relativistic  jet beamed toward  Earth. 

In addition to orientation, intrinsic differences among AGN  also play a role in AGN unification.    For example,  \citet{urry95} conclude their seminal review  with 10 outstanding questions, including one regarding Fanaroff-Riley galaxies \citep[FR;][]{fanaroff74}: ``do FR~Is have broad emission line regions?"  The significance of this question is that if FR~I galaxies lack BELRs, then intrinsic properties might play a prominent role in driving the so-called FR~I/II dichotomy (in which the more powerful FR~II galaxies have edge-brightened radio lobes).  Indications so far are that emission line luminosities from FR~I galaxies are 5--30 times weaker than for FR II galaxies, and FR I galaxies  may also have weaker tori (\citealt{zirbel95, chiaberge99, donato04}; although also see \citealt{cao04, leipski09}).  However, the requisite observations and their interpretation are technically challenging \citep[][]{capetti00_cena, evans06}, and there are still many ongoing ``nature vs.\ nurture" debates.

We can  turn to beamed versions of FR~I  and FR~II galaxies for additional guidance --- i.e., BL Lac objects and flat spectrum radio quasars (FSRQs), respectively, which are collectively  called blazars.  Broad emission lines and sometimes dusty tori are routinely detected in the spectral energy distributions (SEDs) of FSRQs, but less often from BL Lac objects \citep[e.g.][]{abdo10, giommi11}.  However, it is  generally still not observationally clear if BL Lac BELRs and tori appear weak solely due to dilution by jet emission \citep[e.g.,][]{chen11_spitzer, malmrose11}, or rather if they really are intrinsically weak \citep[e.g.,][]{ghisellini11, sbarrato11_ph}.   Constraints on blazar BELRs and tori are furthermore important, as both components are potential sources of seed photons for Comptonized gamma-ray emission, especially for FSRQs and low synchrotron peaked (LSP) BL~Lac objects.  Note, gamma-ray emission from high synchrotron peaked  (HSP)  BL~Lac objects is generally consistent with synchrotron self Compton \citep[e.g.][]{abdo10}.

The preliminary data release of the \textit{Wide-field Infrared Survey Explorer }\citep[\textit{WISE};][]{wright10} recently opened a new multiwavelength window over a large area of the sky.  In this letter, we investigate the tori of  low-luminosity radio galaxies  by examining BL Lac objects detected by \textit{WISE}.   We describe our BL Lac sample in \S \ref{sec:sample}.  In \S \ref{sec:irprop} we describe their IR properties,  and results are discussed in \S \ref{sec:disc}.     All  spectral indices are defined as $f_{\nu} \propto \nu^{-\alpha_{\nu}}$, and we define the broad-band radio to IR spectral index $\alpha_{ri} = -\log(L_r/L_i)/4.25$, where $L_r,$ and $L_i$ are monochromatic luminosities at 5~GHz and 3.4~$\mu$m  rest-frame, respectively.   We adopt the following cosmology:  $H_0=71$~km~s$^{-1}$~Mpc$^{-1}$; $\Omega_M=0.27$; $\Omega_\Lambda=0.73$.

\begin{figure*}
\centering
\includegraphics[scale=0.5]{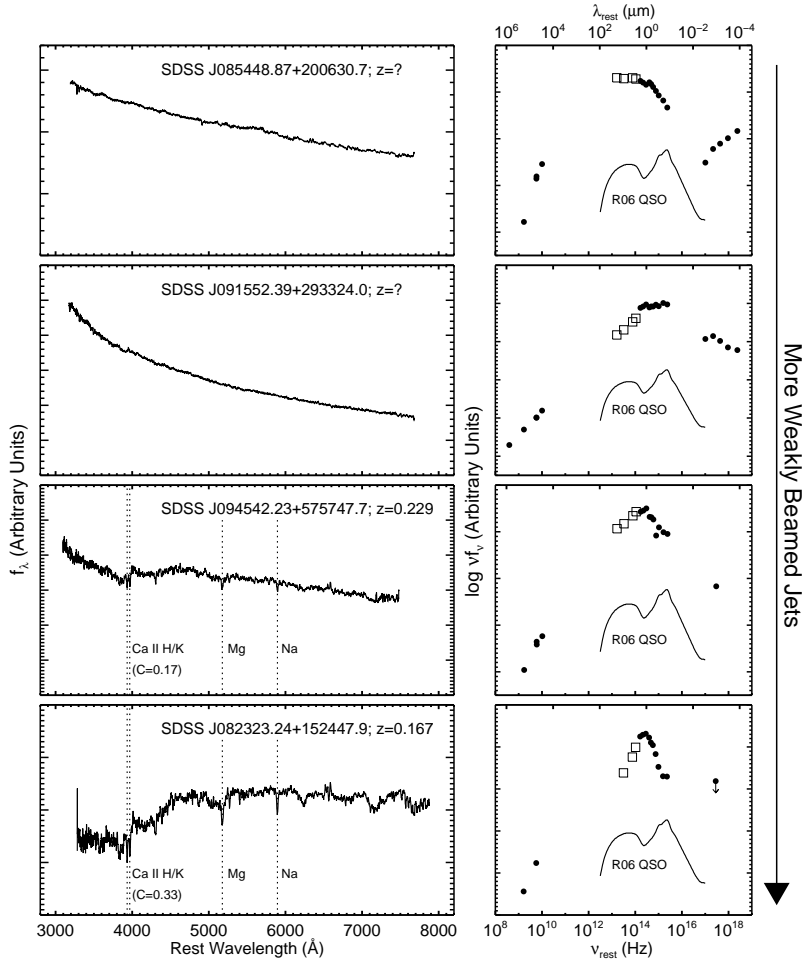}
\caption{Sample SDSS optical spectra for four BL Lac objects (left), and corresponding SEDs (right).  For illustrative purposes, we  assume $z=0.2$ for the top two rows.   \textit{WISE} data points are shown as open squares, and the other SED multiwavelength data are assembled as in \S 4.3.5 of \citet{plotkin11_ph}.  For comparison, we include a typical quasar SED \citep[from][R06]{richards06}, which shows  thermal torus dust emission peaking redward of 1--2$~\mu$m (3.0--1.5$\times10^{14}$~Hz). }
\label{fig:sampspec}
\end{figure*}

\newpage
\section{The BL Lac Sample}
\label{sec:sample}
We start with 590 confirmed radio-loud BL~Lac objects from the catalog of \citet[][hereafter \citetalias{plotkin10}]{plotkin10},\footnote{A subset of radio-quiet BL~Lac candidates from \citetalias{plotkin10} with \textit{WISE} coverage are discussed in \citet{wu11_subm}.  \citet{wu11_subm} show that many are likely low-redshift analogs to weak line quasars (WLQs), which are high-redshift ($z>3$) quasars with unusually weak or missing  BELRs but normal dusty tori \citep[see][]{lane11}. } which were selected  from Sloan Digital Sky Survey spectroscopy \citep[SDSS;][]{york00}.  \citetalias{plotkin10} require no SDSS spectrum to show an emission feature with $REW>5$~\AA, and the \ion{Ca}{2}~H/K break, $C$,  must be smaller than 40\% \citep[e.g.][]{stocke91,marcha96}.  The H/K break  (i.e., the fractional change of continuum flux surrounding 4000~\AA\ rest-frame) quantifies the host galaxy (HG) contribution to each SDSS spectrum.   A spectrum with a more highly beamed jet will show smaller $C$ because the jet is brighter compared to the (unbeamed) HG \citep[see, e.g.][]{landt02}.  Of the $\sim$60\% of BL Lac objects for which \citetalias{plotkin10} can determine redshifts, most (and essentially all at $z<0.5$) are derived via  absorption lines from HG starlight and are thus among the most  weakly beamed \citetalias{plotkin10} BL~Lac objects (see Figure~\ref{fig:sampspec}).

We correlate the  \citetalias{plotkin10} BL~Lac objects  to the preliminary \textit{WISE} data release using a 3$''$ search radius.  We require \textit{WISE} detections with $S/N>3$ in the W1[3.4~$\mu$m], W2[4.6~$\mu$m], and W3[12~$\mu$m]  bands.    Requiring  $S/N>3$ for the W4[22~$\mu$m] band would reduce our sample by at least a factor of two.   We retain 157 objects (see Figure~\ref{fig:sampspec}).   To create comparison samples of normal quasars and early-type galaxies (ETGs; BL Lac objects live perhaps universally in large ellipticals, \citealt{urry00_hstii}), we   also correlate 105,783 spectroscopically confirmed Type~I SDSS  quasars  \citep{schneider10} and 8666 (inactive) early SDSS galaxies   \citep{bernardi03_sample} to \textit{WISE}, applying  the same $S/N$ constraint as above.  All \citetalias{plotkin10} BL~Lac objects have $z<2$, so we also restrict the comparison quasars to $z<2$.   We further consider only quasars with central black hole masses $10^8<M_{bh}/M_{\sun}<10^9$ (masses from \citealt{shen11}), and  ETGs with stellar velocity dispersions $200 < \sigma_{disp} < 320$~km~s$^{-1}$.  The above values of $M_{bh}$ and $\sigma_{disp}$ are typical for BL~Lac HGs \citep[see][]{leontavares11, plotkin11}.  We are left with 13,881 quasars and 747 ETGs.  Throughout, we convert \textit{WISE} magnitudes to flux densities ($f_\nu$) using the the published (iso) zero points.\footnote{http://wise2.ipac.caltech.edu/docs/release/prelim/expsup/figures/ sec4\_3gt4.gif.}

\section{BL Lac Infrared Properties}
\label{sec:irprop} 
\begin{figure}
\centering
\includegraphics[scale=0.4]{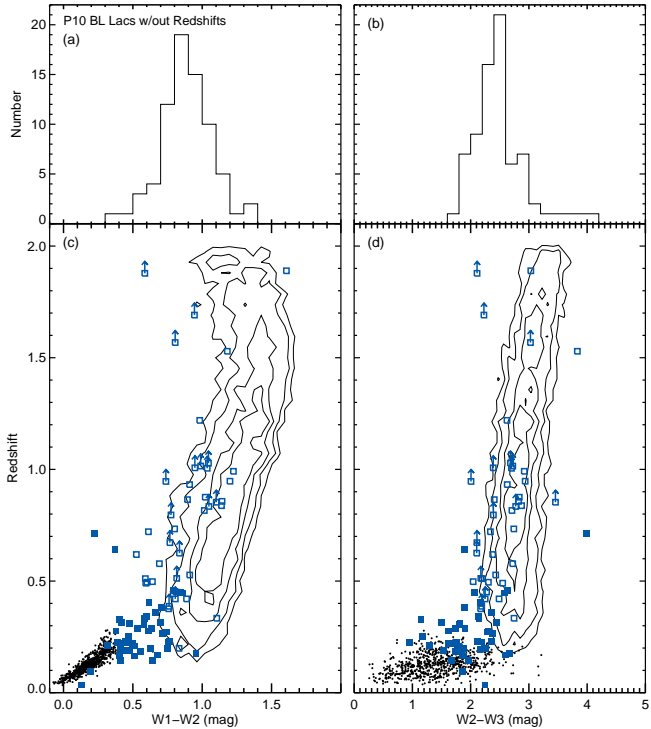}
\caption{Distributions of W1-W2 (a) and W2-W3 (b)  for \citetalias{plotkin10} objects lacking redshifts.  Panels (c) and (d) show redshift vs. IR colors for the remaining  \citetalias{plotkin10} BL~Lac objects (blue squares).  Open symbols indicate less reliable redshifts.  We also show the IR colors of our comparison quasars (contours) and ETGs (filled circles).}
\label{fig:colVz}
\end{figure}

  The  \textit{WISE} colors, W1-W2 and W2-W3,  for the \citetalias{plotkin10} sample and for our comparison quasars and elliptical galaxies are shown in Figure~\ref{fig:colVz} as a function of redshift.    The dusty torus of quasars makes their IR colors redder than for ETGs.   The first-order dependence of the quasar IR colors on redshift is simply due to the  torus  being redshifted through the \textit{WISE} filters.  Although some BL Lac objects have \textit{WISE} colors similar to quasars, BL~Lac objects tend to populate the bluer edge of the quasars' color space.  For most lower-redshift BL Lac objects, this is because of HG contamination (elliptical galaxy SEDs  emit most of their radiation in the near-IR and appear blue in the mid-IR; see Figure~\ref{fig:sampspec}d). However, beamed synchrotron emission should completely dominate the IR emission from the higher redshift objects and those lacking redshifts \citep[see, e.g.][]{padovani06, chen11_spitzer}.   We thus expect any overlap between the IR colors of  highly beamed BL Lac objects and SDSS quasars to be coincidental, since different mechanisms (i.e., synchrotron vs.\ thermal dust emission) are producing their IR flux.
  
\begin{figure*}
\centering
\includegraphics[scale=0.55]{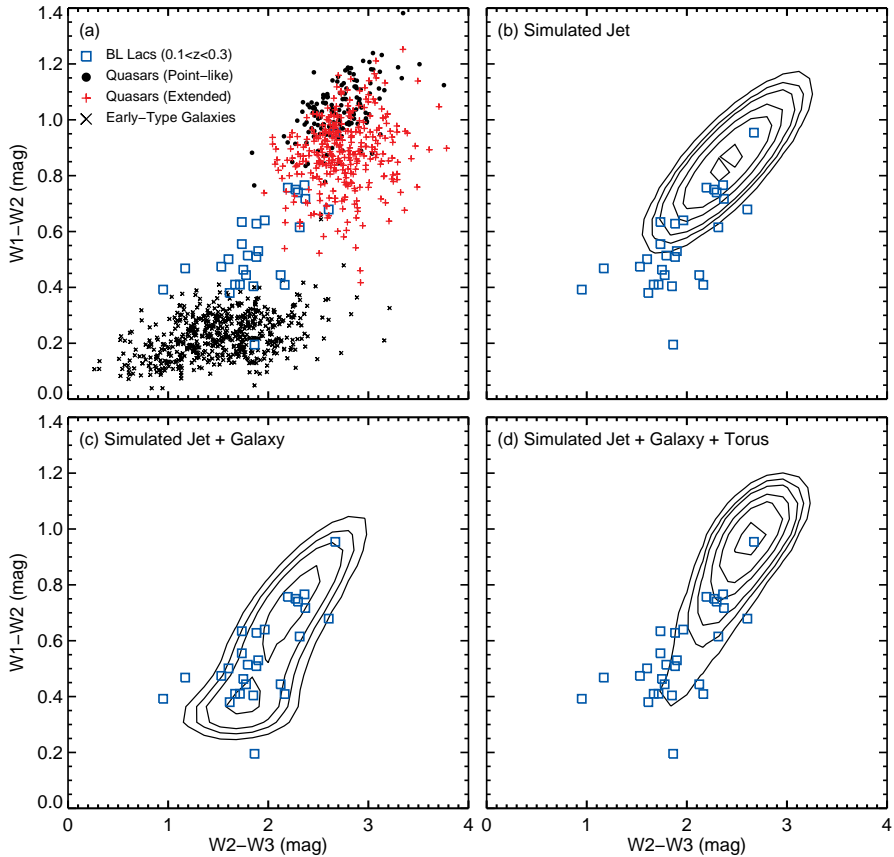}
\caption{(a) Color-color diagram for 28 low-redshift BL Lac objects (blue squares),  150 comparison  quasars with SDSS point-like morphologies (circles), 320 quasars that appear extended in  SDSS imaging,  and 530 ETGs (crosses). In the remaining panels, contours show  expected BL~Lac colors for MC simulations including only a beamed jet (b), a jet and host galaxy (c), and a jet, host galaxy, and torus (d) (see text for details).  Quasars and ETGs are omitted from panels (b)-(d) for clarity.}
\label{fig:wisecolcol}
\end{figure*}

\subsection{Constraints on BL Lac Dusty Tori}
\label{sec:lowzbl}
BL Lac objects  bridge the IR color space between normal galaxies and quasars (Figure~\ref{fig:wisecolcol}a).     One might expect the most weakly beamed BL Lac objects to show contributions from both HG starlight and from thermal dust emission from the torus.  Then, the most weakly beamed BL Lac objects would look more similar to quasars that appear extended in their SDSS images (red plus signs).  However, the weakly beamed BL~Lac objects' IR colors are too blue for a significant contribution from the torus.   In this section we determine if the lack of torus signatures is  simply because even weakly beamed jets (and HGs)  can still  outshine the torus.

All ETGs in our comparison sample have $z<0.3$, so  we restrict ourselves to 28  BL Lac objects with $z<0.3$ in this section.  All 28 objects appear as point sources in their \textit{WISE}  images.  There is thus little concern that the WISE photometry includes different fractions of the total HG flux as a function of redshift.  We also verify that they show no significant correlation between redshift and either \textit{WISE} color.  All 28 objects have $z>0.1$, so, for uniform comparison, we  only consider ETGs and quasars from $0.1<z<0.3$ (530 and 470 objects, respectively).

 There are three emission components that must be modeled in the IR: 1) starlight from the HG; 2) beamed jet synchrotron emission; and 3) dusty torus emission.   To test if the dusty torus is present,   we run Monte Carlo (MC) simulations.   These simulations involve building a set of probability density functions (pdfs; see \S \ref{sec:irhg} --  \S \ref{sec:irtor}), from which we take 10$^6$ random draws to estimate each IR-emitting component's monochromatic luminosity in each \textit{WISE} filter.  From these luminosities  we synthesize the expected \textit{WISE} color space occupied  by BL~Lac populations with and without the  torus.   Throughout we estimate all luminosities at $z=0.2$ to minimize K-correction uncertainties, and we restrict comparison samples to a narrow redshift range near $z=0.2$.   For all pdfs modeled as  (log)-normal distributions below, we first check that their  observed parameter distributions are indeed approximately Gaussian.

\subsubsection{IR Emission from Host Galaxy Starlight}
\label{sec:irhg}
 We consider the observed properties of 164 ETGs with $z=0.22 \pm 0.05$ (the median of our low-redshift BL Lac sample).  We first build a pdf for the monochromatic luminosity  in the W2 filter, assuming a log-normal distribution with $\left<\log L_{\nu}\right>_{hg,w2}=29.82\pm0.14$~erg~s$^{-1}$~Hz$^{-1}$.  Then we make pdfs for HG colors using normal distributions with $\left<W1-W2\right>=0.30\pm0.066$~mag and $\left<W2-W3\right>=1.58\pm0.43$~mag.  Randomly drawing from each distribution allows us to estimate HG monochromatic luminosities in each $WISE$ filter.  This method implicitly accounts for measurement uncertainties. 
 
\subsubsection{Jet Synchrotron in the IR}
\label{sec:irjet}
We begin with an unbeamed 5~GHz radio core luminosity $\left(\nu L_{\nu}\right)_{j, r}$, and we use the broad-band radio-IR spectral index $\alpha_{ri}$ to estimate an unbeamed IR jet luminosity  in the W1 filter, $\left(\nu L_{\nu}\right)_{j,w1}$.  Beamed IR jet luminosity is then calculated as  $\left(\nu L_\nu\right)^\prime_{j,w1}=\left(\nu L_\nu\right)_{j,w1}\delta^{2+\alpha_{wise}}$ \citep[see][]{lind85}, where  $\delta=\left[ \gamma\left(1-\beta \cos\theta\right)\right]^{-1}$ is the Doppler parameter,  $\gamma=(1-\beta^2)^{-0.5}$, $\theta$ is the viewing angle ($\theta = 0$ corresponds to a perfectly aligned jet),  and $\beta$ is the jet velocity normalized to the speed of light.    We then use  $(\nu L_\nu)^\prime_{j,w1}$ and  the local IR spectral index $\alpha_{wise}$ to estimate beamed jet luminosities in the W2 and W3 filters.\footnote{We assume $\alpha_{wise}$ and $\alpha_{ri}$ are not  strongly affected by beaming, which is not strictly true but a reasonable approximation once the jet is highly enough beamed to appear as a BL Lac object.}

To estimate beamed jet luminosities for a random viewing angle, Lorentz factor, and jet power, we randomly draw from the following pdfs: a log-normal distribution for $\left(\nu L_{\nu}\right)_{j, r}$ with $\left<\log\nu L_{\nu}\right>_{j,r}=40.77\pm0.69$~erg~s$^{-1}$ \citep[derived from Equation~2 and Table~4 of][]{wu07}, a normal distribution for $\gamma$ with $\left<\gamma\right>=7\pm0.7$ \citep[e.g., see][]{merloni07}, and a uniformly distributed $\theta$ from 0--40$^\circ$.   For $\alpha_{wise}$, we fit power laws to the W1, W2, and W3 flux densities (in $\log f_\nu-\log\nu$) for the 69  \citetalias{plotkin10} BL~Lac objects lacking redshifts.   Based on those measurements, we then randomly draw $\alpha_{wise}$ from a normal distribution with $\left<\alpha_{wise}\right>=0.60\pm0.43$.  We similarly measure $\alpha_{ri}$ for those 69 BL~Lac objects (assuming $z=0.2$) using the radio fluxes in \citetalias{plotkin10} and the W1 filter.  We then randomly draw $\alpha_{ri}$ from a normal distribution with $\left<\alpha_{ri}\right>=0.40\pm0.11$.   Finally, once we estimate luminosities in each filter as described in the previous paragraph,  we add random noise to each luminosity by assuming  $\sigma_L/L=\pm0.05$, 0.05, and 0.1 in the W1, W2, and W3 filters, respectively (based on typical \textit{WISE} flux measurement uncertainties for our full BL~Lac sample).

\subsubsection{IR Emission from a ``Typical" Obscuring Torus}
\label{sec:irtor}
We employ a realistic coupling between the jet and accretion flow to estimate  IR luminosities from the dusty torus.  First we estimate X-ray luminosity $L_{x}$ (from 2--10~keV) using $\left(\nu L_{\nu}\right)_{j,r}$ from above and the following relations from \citet{merloni07}, based on a sample of low-luminosity AGN with mechanical jet power estimates:
\begin{equation}
\begin{split}
 \log\left(\frac{L_{kin}}{L_{Edd}}\right)  &=  \left(0.49\pm0.07\right)\left(\log\left[\frac{L_{x}}{L_{Edd}}\right]-\log[B/0.2]\right) \\
                      					   & \quad   -(0.78\pm0.36)
\end{split}					     
\label{eq:merxrayjet}
\end{equation} 
and 
\begin{equation}
\noindent \log L_{kin}=\left(0.81\pm0.11\right)\log\left(\nu L_{\nu}\right)_{j,r} + 11.9^{+4.1}_{-4.4},
\label{eq:merradjet}
\end{equation}
\noindent where $L_{kin}$ is the kinetic power of the jet in erg~s$^{-1}$,   $L_{Edd}=1.3\times10^{38}(M/M_{\sun})$~erg~s$^{-1}$ is the Eddington luminosity, and $B$ is fraction of the bolometric luminosity emitted from  2--10~keV.   We note that Equations \ref{eq:merxrayjet} and \ref{eq:merradjet} are consistent with the theoretical scalings for jet dominated accretion flows, $\left(\left[\nu L_{\nu}\right]_{j,r}/L_{edd}\right)\propto(L_{kin}/L_{Edd})^{17/12}$ and $(L_{kin}/L_{Edd})\propto(L_x/L_{Edd})^{0.5}$ \citep[e.g.,][]{falcke95}.  The above scalings are applicable to hard state X-ray binaries and their supermassive analogs,  including BL Lac objects (see, e.g., \citealt{plotkin11_ph}). 

We then assume  X-ray luminosity is a good tracer for the torus' IR luminosity, using the IR/X-ray correlation from \citet{gandhi09}:
\begin{equation}
\begin{split}
 \log\left(\frac{\nu L_{\nu}}{10^{43}}\right)_{12.3\mu m} & =  \left(1.11\pm0.07\right)\log\left(\frac{L_{x}}{10^{43}}\right) \\
                                                                                                   & \quad +\left(0.19\pm0.05\right),
\end{split}
\label{eq:irxray}
\end{equation}
where $(\nu L_{\nu})_{12.3\mu m}$ is the IR luminosity at 12.3 $\mu$m, and  the X-ray luminosity, $L_{x}$, is from 2-10~keV.

Combining Equations \ref{eq:merxrayjet},  \ref{eq:merradjet}, and \ref{eq:irxray} leaves us the following relation for the IR torus emission at 12.3~$\mu$m, assuming an unbeamed radio core luminosity at 5~GHz:
\begin{equation}
\begin{split}
\log\left(\frac{\nu L_{\nu}}{10^{43}}\right)_{12.3\mu m} & =1.84\log\left(\frac{\nu L_\nu}{10^{40}} \right)_{j,r} - 1.16 \log\left(\frac{L_{Edd}}{3.78\cdot10^{46}}\right)  \\
										   & \quad +1.11\log\left(\frac{B}{0.1}\right)+0.43,
\end{split}
\label{eq:radToir}
\end{equation}

\noindent where we assume $B=0.1$ (to conservatively err on the side of underestimating the torus flux).    Finally, we extrapolate torus luminosity to the W2 filter assuming an IR spectral index of 0.15  between 12.3 and 4.6~$\mu$m (estimated at $z=0.2$ from the average \citealt{richards06} quasar SED), and we estimate IR luminosities in the W1 and W3 filters using pdfs based on the \textit{WISE} colors of 150 quasars ($0.1<z<0.3$) with point-like SDSS morphologies.

For the MC simulations, we build the following  additional  pdfs.  For $L_{Edd}$ we assume BL~Lac black hole masses are log-normal with $\left<\log M\right>=8.54\pm0.40~M_{\sun}$ \citep{plotkin11}.   Since Equation~\ref{eq:irxray} relies on several assumptions that propagate non-linearly, we randomly add intrinsic scatter to each estimate of $\left(\log\nu L_{\nu}\right)_{12.3\mu m}$  assuming a generous $\sigma_{int}=\pm1.5$~dex.     For the pdfs for $W1-W2$ and $W2-W3$, we first randomly draw $W1-W2$ from a normal distribution with $\left<W1-W2\right>=1.03\pm0.10$~mag.   Since quasar \textit{WISE} colors are correlated, we then use the best-fit relation  $W2-W3=1.19+1.47(W1-W2)$ to build the $W2-W3$ pdf, and we add random noise  using the \textit{root-mean-square (rms)} scatter about the best-fit regression, $\sigma_{rms}\pm0.23$~mag.

\subsubsection{Lack of observational signatures from the dusty torus}

We perform 10$^6$ MC simulations to estimate luminosities from each IR-emitting component in each \textit{WISE} filter.  Figure~\ref{fig:wisecolcol} shows the synthesized \textit{WISE} colors if only one, two, or all three components are included.  Note, the simulated jets encompass a range of viewing angles $\theta<40^{\circ}$ and Lorentz factors.  \citet{massaro11}  show that blazars fall in a distinct region of IR color space, which they call the `\textit{WISE} blazar strip.'    The most highly beamed \citetalias{plotkin10} BL~Lac objects (i.e., the 69 lacking redshifts) also fall within the strip, and a KS test shows that the W1-W2 and W2-W3 distributions for those 69 BL~Lacs and our synthesized `blazar strip' (panel b) are not significantly different ($p\sim0.4$ and 0.6, respectively).  However, for the 28 weakly beamed objects, a bluer IR component (i.e., the HG) is necessary ($p\sim10^{-11}$ and 10$^{-6}$ from KS tests comparing their W1-W2 and W2-W3 distributions to the synthesized jet).

To explain the SDSS population of weakly beamed BL Lac objects (blue squares), we favor the synthesized colors including the jet and HG (Figure~\ref{fig:wisecolcol}c) over the colors that also include the torus (Figure~\ref{fig:wisecolcol}d).   The simulations including the torus predict that BL~Lac colors  should  be weighted toward redder IR colors, while the simulations with only the jet and galaxy predict  IR colors more consistent with observations.   KS tests show that the  W1-W2 and W2-W3 distributions for 28 BL~Lac objects with $0.1<z<0.3$ and the synthesized colors including the torus are statistically different ($p\sim10^{-11}$ and 10$^{-8}$, respectively), while similar parent distributions are not highly excluded  between the observed colors and the synthesized colors without the torus ($p\sim0.05$ and 0.02).

\section{Discussion and Conclusions}
\label{sec:disc}
We  conclude from the above that our weakly beamed BL Lac objects lack observational signatures of the dusty torus in the mid-IR.  Thus, BL Lac dusty tori appear weaker than the tori of luminous quasars (i.e., one cannot simply scale a normal quasar torus down to BL Lac luminosities).  Another way to express this conclusion is that BL Lac tori have  different properties than the tori of normal quasars.    It is reasonable to extrapolate this  result to highly beamed BL Lac objects, if beaming is primarily a geometric argument.  If one interprets the torus as an extension of the BELR \citep[e.g.,][]{elitzur06}, then our study suggests that BL Lac BELRs are also intrinsically weak.   We stress that this is a statistical conclusion.  The simulations including the torus do predict a small population of weakly beamed BL~Lac objects with \textit{WISE} colors matching the observations, but that population is not large enough to explain the IR colors for the  BL~Lac population as a whole.  However, we cannot exclude the presence of the torus/BELR from every single BL~Lac object.   Indeed,  some BL~Lac objects can show very weak broad emission lines in their optical spectra (especially some LSPs).

One explanation for weaker BL Lac tori is that their physical properties (i.e., torus geometries, dust properties, ionizing SEDs, etc.) are different than for luminous quasars.   Another  interpretation is  the same one invoked for optically dull AGN  and ``naked'' Seyfert galaxies \citep[e.g.,][]{hawkins04, trump11}, as follows.  Consider a scenario where the BELR and torus are fed by a radiatively driven disk wind \cite[e.g.,][]{murray98}.  At sufficiently low accretion rate (below a few percent $L/L_{Edd}$),  the inner region of the accretion disk is replaced by a radiatively inefficient accretion flow (RIAF).  The RIAF is then no longer able to support a sufficient  wind to populate the BELR and torus \citep[e.g.,][]{nicastro00}.  In this scenario, the strength of the BELR and torus is intimately connected to the efficiency of the accretion flow and therefore accretion rate.\footnote{Note, WLQs and their weak BELRs require a different explanation because WLQs show normal dusty tori and are radiatively efficient \citep{lane11}.}  Indeed, such an ``accretion rate dichotomy'' has already been suggested to explain the BL Lac/FSRQ divide \citep[with BL Lac objects fed by RIAFs, e.g.,][]{boettcher02, ghisellini09}.  Such a scheme may also be responsible for the FR~I/II divide \citep[e.g.,][]{ghisellini01, wold07}.  In fact, an important (supporting) point is that the classification of unbeamed radio galaxies based on high-excitation vs.\ low-excitation emission lines (which are likely connected to the accretion flow) is more physically meaningful than the older FR~I/II scheme based on radio morphology/jet power \cite[see, e.g.,][]{jackson99, hardcastle09}.  Furthermore, such a division is naturally expected from  the hysteresis displayed by accreting stellar mass black holes, where changes in X-ray spectral states for individual objects (i.e., at fixed orientations) may  similarly be related to the efficiency of the inner accretion flow.  

 \textit{WISE}  offers a new perspective on the strength of the obscuring torus in low-luminosity radio galaxies.  The \textit{WISE} view of BL Lac objects is  consistent with other (independent) indications of an accretion rate dichotomy among AGN, adding a complementary piece to the puzzle.   Further constraints on such a division in accretion mode   is an important step for better understanding AGN unification.

\acknowledgments
We thank the anonymous referee.  RMP and SM acknowledge support from a Netherlands Organization for Scientific Research (NWO) Vidi Fellowship.  WNB and JW acknowledge support from NASA ADP grant NNX10AC99G.


\end{document}